\documentclass[12pt]{article}
\usepackage{latexsym}
\usepackage{amsmath}
\usepackage{amssymb}
\usepackage{relsize}
\usepackage{geometry}
\def\beqn{\begin{eqnarray}}
\def\eeqn{\end{eqnarray}}

\def\beq{\begin{equation}}
\def\eeq{\end{equation}}
\def\ba{\beq\new\begin{array}{c}}
\def\ea{\end{array}\eeq}

\def\Tr{{\rm Tr}}
\newcommand{\gsim}{\lower.7ex\hbox{$
\;\stackrel{\textstyle>}{\sim}\;$}}
\newcommand{\lsim}{\lower.7ex\hbox{$
\;\stackrel{\textstyle<}{\sim}\;$}}

\newcommand{\ntwo}{${\mathcal N}=2$ }

\newcommand{\ntwot}{${\mathcal N}= \left(2,2\right) $ }
\newcommand{\ntwoo}{${\mathcal N}= \left(0,2\right) $ }
\newcommand{\none}{${\mathcal N}=1$ }
\newcommand{\nonen}{${\mathcal N}=1$}

\newcommand{\p}{\partial}
\newcommand{\wt}{\widetilde}
\newcommand{\ov}{\overline}
\newcommand{\mc}[1]{\mathcal{#1}}
\newcommand{\md}{\mathcal{D}}

\newcommand{\lgr}{\left\lgroup}
\newcommand{\rgr}{\right\rgroup}

\def\slashed#1{\setbox0=\hbox{$#1$}             
   \dimen0=\wd0                                 
   \setbox1=\hbox{/} \dimen1=\wd1               
   \ifdim\dimen0>\dimen1                        
      \rlap{\hbox to \dimen0{\hfil/\hfil}}      
      #1                                        
   \else                                        
      \rlap{\hbox to \dimen1{\hfil$#1$\hfil}}   
      /                                         
   \fi}                                        %

\newcommand{\LN}{\Lambda_\text{SU($N$)}}
\newcommand{\sunu}{{\rm SU($N$) $\times$ U(1) }}

\def\cfl {$\text{SU($N$)}_{\rm C+F}$ }
\def\cfln {$\text{SU($N$)}_{\rm C+F}$}

\newcommand{\aU}{a^{\rm U(1)}}
\newcommand{\aN}{a^\text{SU($N$)}}

\newcommand{\bxir}{\ov{\xi}{}_R}
\newcommand{\bxil}{\ov{\xi}{}_L}
\newcommand{\xir}{\xi_R}
\newcommand{\xil}{\xi_L}

\newcommand{\bzr}{\ov{\zeta}{}_R}
\newcommand{\zr}{\zeta_R}

\newcommand{\nbar}{\ov{n}}

\newcommand{\pts}{\psi_{\rm tr}^0}
\newcommand{\ptN}{\psi_{\rm tr}^N}
\newcommand{\cts}{\chi_{\rm tr}^0}
\newcommand{\ctN}{\chi_{\rm tr}^N}
\newcommand{\tts}{\vartheta_{\rm tr}^0}
\newcommand{\ttN}{\vartheta_{\rm tr}^N}
\newcommand{\rts}{\rho_{\rm tr}^0}
\newcommand{\rtN}{\rho_{\rm tr}^N}

\newcommand{\tor}{\vartheta_{\rm or}}
\newcommand{\eor}{\eta_{\rm or}}
\newcommand{\kor}{\kappa_{\rm or}}
\newcommand{\sor}{\sigma_{\rm or}}
\newcommand{\por}{\psi_{\rm or}}
\newcommand{\cor}{\chi_{\rm or}}
\newcommand{\uor}{\upsilon_{\rm or}}
\newcommand{\oor}{\omega_{\rm or}}

\newcommand{\CPC}{CP($N-1$)$\times$C }
\newcommand{\CPCn}{CP($N-1$)$\times$C}

\newcommand{\tgamma}{\wt{\gamma}}

\begin{document}

\begin{titlepage}

\begin{flushright}
FTPI-MINN-09/10, UMN-TH-2740/09\\
March 4, 2009
\end{flushright}

\begin{center}

{\Large \bf   Description of the Heterotic String Solutions
\\[2mm]
 in the \boldmath{$M$} Model }
\end{center}

\begin{center}
{\bf P. A. Bolokhov$^{a,b}$, M.~Shifman$^{c}$, and \bf A.~Yung$^{c,d}$}
\end {center}
\vspace{0.3cm}
\begin{center}

$^a${\it Physics and Astronomy Department, University of Pittsburgh, Pittsburgh, Pennsylvania, 15260, USA}\\
$^b${\it Theoretical Physics Department, St.Petersburg State University, Ulyanovskaya~1, 
	 Peterhof, St.Petersburg, 198504, Russia}\\
$^c${\it  William I. Fine Theoretical Physics Institute,
University of Minnesota,
Minneapolis, MN 55455, USA}\\
$^{d}${\it Petersburg Nuclear Physics Institute, Gatchina, St. Petersburg
188300, Russia}\\

\end{center}

\begin{abstract}

We continue the  study of heterotic non-Abelian BPS-saturated flux tubes (strings).
Previously, such solutions were obtained  in  U($N$) gauge theories:
 \ntwo super\-symmetric QCD deformed by
superpotential terms $\mu{\mathcal A}^2$ breaking
\ntwo supersymmetry down to \none$\!\!$. 
In these models one cannot consider the limit $\mu\to\infty$
which would eliminate adjoint fields: the bulk theory develops a Higgs
branch; the emergence of massless particles in the bulk
precludes one from taking the  limit $\mu\to\infty$.
This drawback is absent in the $M$ model (hep-th/0701040)
where the matter sector includes additional ``meson"
fields $M$ introduced in a special way.
We generalize our previous results 
to the $M$ model, derive the
 heterotic string (the 
 string world-sheet theory is
 a heterotic \ntwoo 
sigma model, with the CP$(N-1)$ target space for bosonic fields and an extra 
right-handed fermion coupled to the fermion fields of the
\ntwot CP$(N-1)$ model),
 and then explicitly obtain all relevant zero modes.
This allows us to relate parameters of the
microscopic $M$ model to those of  the world-sheet theory.  
The limit $\mu\to\infty$ is perfectly smooth.
Thus, the full-blown and fully analyzed heterotic string emerges, for the first time, in
the \none theory with no adjoint fields. The fate of the confined monopoles is discussed.

\end{abstract}

\end{titlepage}

\section{Introduction}

Non-Abelian BPS-saturated flux tubes were discovered \cite{HT1,ABEKY} and studied 
\cite{SYmon,Tong,HT2} in \ntwo super\-symmetric QCD with the gauge group
U$(N)$, the Fayet--Iliopoulos (FI) term,
and $N$ flavors ($N$ hypermultiplets in the fundamental representation),
for reviews see \cite{Trev,Jrev,SYrev,Trev2}.
If \ntwo supersymmetry is maintained in the bulk,
the low-energy theory on the
string world sheet is split into two disconnected parts:
a free theory for (super)translational moduli and a nontrivial part, a theory of
interacting (super)orientational moduli described by \ntwo supersymmetric CP$(N-1)$ 
sigma model.  The above splitting of the moduli space
 is completely predetermined by the fact
that the basic bulk theory has eight supercharges, and the string under consideration
is 1/2 BPS (classically). 

If \ntwo bulk theory is deformed by mass terms $\mu{\mathcal A}^2$ of the adjoint fields,
breaking \ntwo down to \nonen, the situation drastically changes:
two of the four former supertranslational modes
become coupled to two superorientational modes \cite{Edalati}.
As a result, the world sheet theory is deformed too.
Instead of the well-studied ${\mathcal N}=(2,2)$ CP$(N-1)$ model we now have
a heterotic \ntwoo 
sigma model, with the CP$(N-1)$ target space for bosonic fields and an extra 
right-handed fermion which couples to the fermion fields of the
CP$(N-1)$ model in a special way. In the previous works \cite{SYhet,BSYhet}
the heterotic world-sheet model was derived from the microscopic theory by
a direct calculation of all relevant zero modes.
This allowed us 
to relate the  heterotic \ntwoo 
sigma model parameters with those of the bulk theory.

The task we addressed was moving away from \ntwo$\!\!$, towards \none$\!\!$.
In particular, it is highly desirable to get rid of all adjoint fields
inherent to \ntwo models.  If we were able to tend $\mu\to\infty$ this goal would be achieved,
all adjoint fields would become infinitely heavy and could be eliminated. Unfortunately,
simultaneously with increasing the masses of the adjoint fields the bulk theory develops 
a Higgs branch, and massless (light) moduli fields come with it. The string swells,
and all approximations fail at $\mu >\mu_*$ where $\mu_*$ is a critical value,
\beq
\mu_* \sim \frac{\xi}{\Lambda_\sigma}\,,
\label{critmu}
\eeq
$\xi$ is the FI coefficient and $\Lambda_\sigma$ is the dynamical scale
of the world-sheet sigma model.
Although $\mu_*$ can be made large, there are crucial questions which cannot be addressed
under the constraint $\mu \ll \mu_*$. One of them is the fate of the kinks in the heterotic
CP$(N-1)$ model which, from the bulk standpoint, represent confined monopoles.
At $\mu \ll \mu_*$ the world-sheet theory has $N$ degenerate (albeit quantum-mechanically nonsupersymmetric) vacua
which are well defined. Correspondingly, the kink masses are well-defined too;
in fact, they were calculated \cite{SYhet2} in the large-$N$ limit. In a formal limit $\mu\to\infty$
the above degenerate vacua coalesce. Physics of the kinks becomes obscure.

To avoid this problem an $M$ model was designed \cite{GSYmmodel}. Besides the fields present in the
\none deformation of the basic \ntwo bulk theory, the $M$ model includes $N^2$
``mesonic" superfields, which break \ntwo right from the start. The $M$ model is characterized by one
extra interaction constant $h$. 
It was demonstrated \cite{GSYmmodel} that at finite $h$ the limit $\mu\to\infty$ becomes smooth.
Therefore, one can completely eliminate the adjoint fields.
The solitonic flux tube solution (BPS-saturated at the classical level) persists.  
Our task in this paper is to derive the world-sheet theory for these strings (in the limit of
small $h$ and $\mu\to\infty$). We prove that it is the same heterotic
CP$(N-1)$ model, with specific relations between constants of this model and those of the bulk theory.
To obtain these relations we determine all relevant zero modes for the $M$-model flux-tube solution.


The paper is organized as follows. In Sect.~2 we review $M$ model, string solutions and 
the bosonic part of the world-sheet theory on the non-Abelian string. In Sect.~3 we calculate
fermionic supertranslational and superorientational modes of the string and derive
the fermionic part of the world-sheet theory. Our derivation shows that the world-sheet theory is
the  heterotic \ntwoo supersymmetric CP$(N-1)$ model. In Sect.~4 we discuss the fate of
the  bulk monopoles
confined on the string in the limit of large $\mu$. Section~5 contains our brief conclusions.
Our notation is summarized  in Appendix of Ref.~\cite{BSYhet}.


%
%

%
%
\section{The {\boldmath $M$} Model}
\setcounter{equation}{0}

In this section we describe how non-Abelian strings emerge in the $M$ Model.
Our discussion here parallels that of \cite{GSYmmodel} where more details are
given, and we review only the most essential points.
The theory we start with is  \ntwo Supersymmetric Quantum Chromodynamics (SQCD) 
with the gauge group \sunu  and \ntwo supersymmetry
explicitly broken to \none by the following deformations:
\beq
\label{none}
	\delta\mc{W}_{\mc{N}=1} ~~=~~ \sqrt{2N}\,\mu_1\, \left( \mc{A}^{\rm U(1)} \right)^2 
				~+~ \frac{\mu_2}{2}\, (\mc{A}^a)^2
				~+~ \Tr\, M\, \wt{Q}\, Q\,.
\eeq
Here the first two terms break supersymmetry by giving masses to the adjoint 
supermultiplets $ \mc{A}^{\rm U(1)} $ and $ \mc{A}^\text{SU($N$)} $, 
while $ M $ breaks supersymmetry by coupling to the quark fields. In fact,
$ M $ is the superfield extension of the quark mass matrix $ m_A^B $. It promotes 
$ m_A^B $ to a dynamical chiral superfield,
\[
	\delta S_{M \rm kin} ~~=~~ \int\, d^4x\, d^2\theta\, d^2\ov{\theta} \;
					\frac{2}{h}\, \Tr\, \ov{M}\, M \,,
\]
where $ h $ is a dimensionless coupling constant. 

\vspace{1mm}

Why do we introduce $\mu{\mathcal A}^2$ deformations?
The reason behind the introduction of the masses $ \mu_1 $ and $ \mu_2 $ is 
the desire to make the adjoint fields
heavy and exclude them from low-energy physics.
The role of the superfield $ M $ is to lift the Higgs branch which appears when
the adjoints are integrated out. If the coupling constant $h=0$, the $M$ field is frozen at an arbitrary  constant 
and returns to its original status  of the quark mass matrix. This does not break \ntwo supersymmetry. However,
once $h\neq 0$, the coupling to $M$ becomes a deformation which breaks \ntwo supersymmetry
down to \none (in conjunction  with nonvanishing parameters $\mu_1$ and $\mu_2$).

\vspace{1mm}

The bosonic part of the theory is
\begin{align}
\label{theory}
	S_{\rm bos} ~~=~~ & \int d^4 x 
		\lgr
			\frac{1}{2g_2^2}\Tr \left(F_{\mu\nu}^\text{SU($N$)}\right)^2  ~+~
			\frac{1}{g_1^2} \left(F_{\mu\nu}^{\rm U(1)}\right)^2 ~+~ 
			\right. 
			\\[3mm]
\notag
		&
			\phantom{\int d^4 x \lgr\right.\,}
			\frac{2}{g_2^2}\Tr \left|\nabla_\mu \aN \right|^2   ~+~
			\frac{4}{g_1^2} \left|\p_\mu \aU \right|^2
			~+~
			\left| \nabla_\mu q^A \right|^2 ~+~ \left|\nabla_\mu \ov{\wt{q}}{}^A \right|^2 
			~+~
			\\[3mm]
\notag
		&
			\phantom{\int d^4 x \lgr\right.\,}
		\left.
			\frac{1}{h}\, \left| \p_\mu M^0 \right|^2  ~+~
			\frac{1}{h}\, \left| \p_\mu M^a \right|^2 ~+~
			V(q^A, \wt{q}{}_A, a^a, \aU, M^0, M^a)
		\rgr .
\end{align}
Here $ \nabla_\mu $ is the covariant derivative in the appropriate representation
\begin{align*}
	\nabla_\mu^{\rm adj} & ~~=~~ \p_\mu  ~-~ i\, [ A_\mu^a T^a, \;\cdot\;]~, \\[3mm]
	\nabla_\mu^{\rm fund} & ~~=~~ \p_\mu ~-~ i\,A^{\rm U(1)}_\mu ~-~ i\, A_\mu^a T^a\,.
\end{align*}
The vector fields $A_{\mu}$ and complex scalars $a$ belong to gauge multiplets of the U(1) and SU$(N)$ sectors, respectively, while $q^{kA}$ and  $\tilde{q}_{Ak}$ denote squarks, $k=1,...,N$ and $A=1,...,N_f$ are the
color and flavor indices, respectively. In this paper we consider only the case $N_f=N$.

\vspace{1mm}

	The matrix superfield $ M^A_B $ is conveniently decomposed as 
\[
	M^A_B ~~=~~ \frac{1}{2}\,\delta^A_B\, M^0  ~+~ (T^a)^A_B\, M^a\,.
\]
Assembling results of Refs. \cite{SYhet,GSYmmodel} (see also \cite{SYrev}) one can readily see that the  potential of the theory \eqref{theory} takes the form
\begin{align}
\notag
	& V(q^A, \wt{q}{}_A, a^a, \aU, M^0, M^a) ~~=~~ 
	\\[3mm]
\notag
	&\qquad\quad ~~=~~
			\frac{g_2^2}{2} \left( \frac{1}{g_2^2}\,f^{abc}\ov{a}{}^b a^c 
				~+~ \Tr\, \ov{q}\, T^a q 
				~-~ \Tr\, \wt{q}\, T^a \ov{\wt{q}} \right)^2 
	\\[3mm]
\label{V}
	&\qquad\quad ~~+~~
		\frac{g_1^2}{8}\, (\Tr\, \ov{q} q ~-~ \Tr\, \wt{q} \ov{\wt{q}} ~-~ N\xi )^2
	\\[3mm]
\notag
	&\qquad\quad ~~+~~
		2g_2^2\, \Bigl| \Tr\, \wt{q}\,T^a q ~+~ 
			\frac{1}{\sqrt{2}}\, \frac{\p\mc{W}_{\mc{N}=1}}{\p a^a} \Bigr|^2
	~+~
	\frac{g_1^2}{2}\, \Bigl| \Tr\, \wt{q} q ~+~ 
			\frac{1}{\sqrt{2}}\, \frac{\p\mc{W}_{\mc{N}=1}}{\p\aU} \Bigr|^2
	\\[3mm]
\notag
	&\qquad\quad ~~+~~
	2\, \Tr \Biggl\{  
		\left| \left( \aU ~+~ a^a T^a \right) q ~+~
			\frac{1}{\sqrt{2}}\, q\left( \frac{M^0}{2} \;+\; M^a\,T^a \right) \right|^2 
	\\[3mm]
\notag
	&\phantom{\qquad\quad ~~+~~ 2 } ~+~
		\left| \left( \aU ~+~ a^a T^a \right) \ov{\wt{q}} ~+~
			\frac{1}{\sqrt{2}}\, \ov{\wt{q}}\left( \frac{M^0}{2} \;+\; M^a\,T^a \right) \right|^2 
			\Biggr\}
	\\[3mm]
\notag
	&\qquad\quad ~~+~~
	\frac{h}{4}\, \left| \Tr\, \wt{q}q \right|^2  ~+~ h\,\left| \Tr\, q\,T^a\,\wt{q} \,\right|^2	
	~.
\end{align}
	This potential is a sum of $ F $ and $ D $ terms, in particular the last two terms in \eqref{V}
	are the $ F $ terms of the  $ M $ field.
	We also introduced the Fayet--Iliopoulos $ D $ term in the third line, with the real (and positive)
	parameter $ \xi $.
	The \ntwo supersymmetry is broken by parameters $ \mu_1 $, $ \mu_2 $ and $ h $ via $ \mc{W}_{\mc{N}=1} $, while
	the FI term does not break \ntwo supersymmetry \cite{HSZ,VY}.

	A nonvanishing $ \xi $ in the potential triggers condensation of quarks and spontaneous breaking 
	of the gauge symmetry. 
	The vacuum expectation values (VEVs) of the quarks can be chosen in the form
\begin{align}
\notag
&
	\langle q^{kA} \rangle ~=~ \sqrt{\xi} 
		\begin{pmatrix}
			 1  &   0  &  ... \\
			... &  ... &  ... \\
			... &   0  &  1 
		\end{pmatrix} ~,
	\qquad\qquad 
	\langle \ov{\wt{q}}{}^{kA} \rangle ~=~ 0~,
	\\
\label{qvev}
&
	\qquad\qquad  k~=~ 1,...\, N~, \qquad  A ~=~ 1,...\, N~,
\end{align}
i.e.	the so-called color-flavor locked form.
	The adjoint VEVs have to vanish classically,
\beq
\label{avev}
	\langle \aN \rangle  ~~=~~ 0~, \qquad\qquad  \langle \aU \rangle ~~=~~ 0~,
\eeq
	together with the VEVs of the $ M $ field,
\beq
\label{Mvev}
	\langle M^a \rangle ~~=~~ 0~, \qquad\qquad  \langle M^0 \rangle ~~=~~ 0~.
\eeq

	Despite the full Higgsing of the gauge symmetry, the VEVs \eqref{qvev},
	\eqref{avev}, and \eqref{Mvev} leave a global diagonal \cfl symmetry unbroken,
\beq
\label{c+f}
	q ~~\to~~ UqU^{-1}\,, \qquad \aN ~~\to~~ U \aN U^{-1}\,, \qquad
		M ~~\to~~ UMU^{-1}\,.
\eeq
	In what follows, we will be interested in the limit of very large $ \mu_1 $, $ \mu_2 $.
	It appears that the VEV structure \eqref{qvev}, \eqref{avev} and \eqref{Mvev} does not
	depend on the supersymmetry breaking parameters, owing to the fact that 
	the adjoint fields vanish in the vacuum, see Eq.~\eqref{avev}.
	In particular, the VEVs will retain the same form up to very large $ \mu $. 

	To allow the theory to be treated semiclassically, we arrange it to be at weak
	coupling, by separating the dynamical scale of SU($N$) from the scale of the gauge
	symmetry breaking $ \xi $ as follows:
\[
 \sqrt{\xi} ~\gg~ \LN \,.
\]

	The perturbative spectrum was discussed in detail in \cite{GSYmmodel}, and we will only
	concentrate on the limit of large $ \mu $. 
	Regardless of $ \mu $, the gauge bosons acquire mass,
\beq
\label{phmass}
	m_{\rm ph} ~~=~~ g_1\, \sqrt{\frac{N}{2}\,\xi}
\eeq
	for the U(1) gauge boson (``photon'') and
\beq
\label{wmass}
	m_W ~~=~~ g_2\, \sqrt{\xi}
\eeq
	for the SU($N$) bosons.

	The scalar bosons line up in the following hierarchy of scales.
	The heaviest bosons, in the $ \mu_i \gg \sqrt{\xi} $ limit, have the masses
\begin{align}
\label{amass}
\notag
	m_{\rm U(1)}^{\rm(largest)} & ~~=~~ \sqrt{\frac{N}{2}}\, g_1^2\mu_1 \,, \\
	m_\text{SU($N$)}^{\rm(largest)} & ~~=~~ \phantom{\sqrt{\frac{N}{2}}\, }
						g_2^2\mu_2\,,
\end{align}
	with the first mass carried by two degenerate states, while
	the second mass is carried by $ 2 (N^2 - 1) $ states.
	These are the masses of heavy adjoint scalars $ \aU $ and $ \aN $.
The low-energy bulk theory spectrum consists of light states with the masses
\begin{align}
\label{U1mass}
\notag
	m_\text{U(1)}^{(1)}  ~~=~~ \sqrt{\frac{h N\xi}{4}}
		\left\{  1 ~+~ \frac{\sqrt{\xi}}{2g_1\mu_1}\,\sqrt{\gamma_1 (\gamma_1 + 1)} 
				~+~ \cdots \right\}\,,	
	\\
	m_\text{U(1)}^{(2)}  ~~=~~ \sqrt{\frac{h N\xi}{4}}
		\left\{  1 ~-~ \frac{\sqrt{\xi}}{2g_1\mu_1}\,\sqrt{\gamma_1 (\gamma_1 + 1)}
				~+~ \cdots \right\}\,,
\end{align}
	with two states for each, and of states with the masses
\begin{align}
\label{SUNmass}
\notag
	m_\text{SU($N$)}^{(1)} ~~=~~ \sqrt{\frac{h\xi}{2}}
		\left\{ 1 ~+~ \frac{\sqrt{\xi}}{2 g_2\mu_2}\,\sqrt{\gamma_2 (\gamma_2 + 1)}
				~+~ \cdots \right\}\,,
	\\
	m_\text{SU($N$)}^{(2)} ~~=~~ \sqrt{\frac{h\xi}{2}}
		\left\{ 1 ~-~ \frac{\sqrt{\xi}}{2 g_2\mu_2}\,\sqrt{\gamma_2 (\gamma_2 + 1)}
				~+~ \cdots \right\}\,,
\end{align}
	with $ 2 (N^2 - 1) $ degenerate states for each of the values.

	At non-zero $ h $ there are no massless states in the bulk theory, even
	if $ \mu_i \to \infty $.
	One can integrate out the heavy adjoint fields, obtaining
\begin{align}
\label{mmodel}
\notag
	S_{\rm bos}^\text{$M$ model} ~=~ & \int d^4 x 
		\biggl\lgroup
			\frac{1}{2g_2^2}\Tr \left(F_{\mu\nu}^\text{SU($N$)}\right)^2  ~+~
			\frac{1}{g_1^2} \left(F_{\mu\nu}^{\rm U(1)}\right)^2 ~+~ 
			\Tr\, \left| \nabla_\mu q \right|^2 ~+~ \Tr\,\left|\nabla_\mu \ov{\wt{q}}\, \right|^2 
			\\[2mm]
		&
\notag
			\phantom{\int d^4 x}
			~+~
			\frac{1}{h}\, \left| \p_\mu M^0 \right|^2  ~+~
			\frac{1}{h}\, \left| \p_\mu M^a \right|^2 ~+~
			\frac{g_2^2}{2} ( \Tr\, \ov{q}\, T^a q 
					~-~ \Tr\, \wt{q}\, T^a \ov{\wt{q}} )^2 
			\\[2mm]
\notag
		&
			\phantom{\int d^4 x}
			~+~
			\frac{g_1^2}{8}\, (\Tr\, \ov{q} q ~-~ \Tr\, \wt{q} \ov{\wt{q}} ~-~ N\xi )^2
			~+~
			\Tr\, |q\,M|^2 ~+~ \Tr\, |\ov{\wt{q}}\, M |^2
			\\[2mm]
		&
			\phantom{\int d^4 x}
			~+~
			\frac{h}{4}\, \left| \Tr\, \wt{q}q \right|^2  ~+~ 
			h\,\left| \Tr\, q\,T^a\,\wt{q} \,\right|^2
			\biggr\rgroup\,.	
\end{align}
	The vacuum of this theory is given in Eqs.~\eqref{qvev} and \eqref{Mvev}.
	The perturbative excitations consist of the \none gauge multiplets with masses
	\eqref{phmass} and \eqref{wmass}, and of chiral multiplets with masses determined by the 
	leading terms in Eqs.~\eqref{U1mass} and \eqref{SUNmass}.
	The scale of the \none theory is related to that of the original \ntwo theory as follows:
\beq
	\Lambda_{\mc{N}=1}^{2N} ~~=~~ \mu_2^N\, \Lambda^N_\text{SU($N$)}\,.
\label{Lambda}
\eeq
	By taking the FI parameter large enough, 
	$ g_2\sqrt{\xi} \gg \Lambda_{\mc{N}=1} $,
	we ensure the \none theory (the $M$ model) is at
	weak coupling.
	
	The theory \eqref{mmodel} admits the existence of non-Abelian strings, the presence of which can be traced
	from the \ntwo theory \eqref{theory}.
	A $ Z_N $ string can be written in terms of the profile functions \cite{ABEKY,GSYmmodel}
\begin{align}
\notag
	q   ~~=~~  &
		\lgr \begin{matrix}
			\phi_2(r) & 0     & \dots      & 0      \\
			\dots     & \dots & \dots      & \dots  \\
			0         & \dots & \phi_2(r)  & 0      \\
			0         &  0    & \dots      & e^{i\alpha}\phi_1(r) 
		     \end{matrix} \rgr ,
		\qquad \wt{q} ~~=~~ 0\,,
	\\
\label{znstr}
	\\[-0.7cm]
\notag
	A_i^\text{SU($N$)}  ~~=~~
		\frac{1}{N} & \lgr \begin{matrix}
        			    	1       &   \dots   &  0       &   0   \\
        				\dots   &   \dots   &  \dots   & \dots \\
        				0       &   \dots   &  1       &   0   \\
        				0       &     0     &  \dots   & - (N-1) 
	   		         \end{matrix} \rgr
		(\p_i\alpha)\bigl( -1 ~+~ f_{N}(r) \bigr)\,,
	\\
\notag
	A_i^{\rm U(1)}  ~~=~~ & \frac{1}{N}(\p_i\alpha)\lgr 1 ~-~ f(r) \rgr \cdot \mathlarger{\mathbf{1}}\,,
	\qquad 
	A_0^{\rm U(1)} ~=~ A_0^\text{SU($N$)} ~=~ 0\,,
	\\
\notag
	\aU ~~=~~ & \aN ~~=~~ M^0 ~~=~~ M^a ~~=~~ 0\,,
\end{align}
	where $ r $ and $ \alpha $ are the polar coordinates in the plane orthogonal to the string, while 
the	index $ i = 1,~ 2 $ labels the Cartesian coordinates in this plane.
	The quark profile functions $ \phi_1(r) $, $ \phi_2(r) $, and the gauge profile functions 
	$ f(r) $ and $ f_N(r) $ obey a system of first-order differential equations
\begin{align}
\notag
&	\p_r\, \phi_1(r) ~-~ \frac{1}{Nr}\, \lgr f(r) ~+~ (N-1)f_{N}(r) \rgr \phi_1(r) ~~=~~ 0, \\
\notag
&	\p_r\, \phi_2(r) ~-~ \frac{1}{Nr}\, \lgr f(r) ~-~ f_{N}(r) \rgr \phi_2(r) ~~=~~ 0 ,\\
\label{foes}
&	\p_r\, f(r) ~-~ r\, \frac{N g_1^2}{4} \lgr (N-1)\phi_2(r)^2 ~+~ \phi_1(r)^2 ~-~ N\xi \rgr ~~=~~ 0 , \\
\notag
&	\p_r\, f_{N}(r)  ~-~  r\, \frac{g_2^2}{2} \lgr \phi_1(r)^2 ~-~ \phi_2(r)^2 \rgr ~~=~~ 0~,
\end{align}
	with the boundary conditions
\begin{align}
\label{boundary}
	\phi_1(0) & ~~=~~  0\text,                   & \phi_2(0) & ~~\neq~~ 0\text,  &
	\phi_1(\infty) & ~~=~~ \sqrt{\xi} \text,     & \phi_2(\infty) & ~~=~~ \sqrt{\xi}\text, \\
\notag
	f_{N}(0) & ~~=~~ 1\text,                   & f(0) & ~~=~~ 1\text,   &
	f_{N}(\infty) & ~~=~~ 0 \text,            &  f(\infty) & ~~=~~ 0\text.
\end{align}
	The tension of the $ Z_N $ string \eqref{znstr} is 
\[
	T_1  ~~=~~ 2\pi\xi~.
\]
	
	Besides the position of the center of the string $ x_0 $, a genuine non-Abelian string 
	also possesses collective coordinates in the group space \cfln, which determine the orientation 
	of the string in the group.	
	The solution \eqref{znstr} breaks \cfl down to SU($N-1$) $\times$ U(1).
	Therefore, the space of the orientational coordinates is given by the coset
\beq
\label{modulispace}
	\frac{\text{SU($N$)}}
            {\text{SU($N-1$)} \times {\rm U(1)}}         ~~\sim~~  \text{CP($N-1$)}\,.
\eeq
	A general non-Abelian string solution can be obtained from Eqs.~\eqref{znstr} by applying
	a \cfl rotation $ U $, namely,
\begin{align}
\notag
	q ~~=~~ & U\, \lgr \begin{matrix}
			   	\phi_2(r)  & 0  & \ldots & 0 \\
				\ldots  &  \ldots & \ldots & \ldots \\
				0  & \ldots      & \phi_2(r) &  0 \\
				0  & 0           & \ldots  &  \phi_1(r) 
			   \end{matrix}        \rgr     
			U^{-1} \,,
		\qquad \wt{q} ~~=~~ 0\,,
		\\[2mm]
\label{nastr}
	A_i^\text{SU($N$)} ~~=~~ \frac{1}{N}\, &\, U\, \lgr \begin{matrix}
					          	1  & \ldots & 0 & 0 \\
						  	\ldots & \ldots & \ldots & \ldots \\
							0  & \ldots  & 1  &  0 \\
							0  & 0   & \ldots  &  - (N-1) 
					         \end{matrix} \rgr  \, U^{-1} (\p_i \alpha)\, f_{N}(r)\,,  \\[2mm]
\notag
	A_i^{\rm U(1)} ~~=~~ & -\,\frac{1}{N}\, (\p_i \alpha)\, f(r) \cdot \mathlarger{\mathbf{1}}\,,
	\qquad\qquad
			A_0^{\rm U(1)} ~~=~~ A_0^\text{SU($N$)} ~~=~~ 0\,,
	\\
\notag
	\aU ~~=~~ & \aN ~~=~~ M^0 ~~=~~ M^a ~~=~~ 0\,,
\end{align}
	where we have passed to the singular gauge in which the quark field does not wind, while the
	gauge field winds around the origin.

	The bosonic string solution \eqref{nastr} at the classical level
	does not involve $ \wt{q} $, the adjoint fields $ a $,
	or the $ M $ fields;  thus it is independent of the supersymmetry
	breaking parameters.
	In particular, this solution will retain its form when $ \mu_2 $ is taken very large and the 
	adjoints are integrated out.
	Therefore, this solution will still be present in the $M$ model \eqref{mmodel}, 
\begin{align}
\notag
	q & ~~=~~ 
		\phi_2 ~+~ n\nbar\, \bigl( \phi_1 ~-~ \phi_2 \bigr) \,,
	\\[2mm]
%
\label{str}
	A_i^\text{SU($N$)} & ~~=~~ \varepsilon_{ij}\, \frac{x^i}{r^2}\, f_{N}(r)
				\lgr n\nbar ~-~ 1/N \rgr,
	\\[2mm]
\notag
	A_i^{\rm U(1)} & ~~=~~ \frac{1}{N}\varepsilon_{ij}\, \frac{x^i}{r^2}\, f(r)~, 
	\\[2mm]
\notag
	\wt{q} & ~~=~~ M^0 ~~=~~ M^a ~~=~~ 0\,.
\end{align}
	We have introduced here the orientational collective coordinates $ n^l $, which
	parametrize the rotation matrix $ U $ as follows
\beq
\label{n}
	\frac{1}{N}\, U \, \lgr \begin{matrix}
				  1  & \ldots & 0 & 0 \\
				  \ldots & \ldots & \ldots & \ldots \\
				  0 & \ldots & 1 & 0  \\
				  0 & 0 & \ldots & -(N-1) 
				\end{matrix} \rgr
			U^{-1}  
	~~=~~
	-\, n^i\,\ov{n}{}_l  ~~+~~ \frac{1}{N}\cdot{\mathlarger{\mathbf{1}}}{}^i_{~l} ~,
\eeq
	where we deploy matrix notation on the left-hand side.
	The coordinates $ n^l $ ($ l = 1, ..., N $) form a complex vector in the fundamental representation
	of SU($N$) and live in the CP($N-1$) space, {\it i.e.}
\beq
\label{unitvec}
		\ov{n}{}_l \cdot n^l ~~=~~ 1,
\eeq
	and one common complex phase of $ n^l $ can be gauged away ({\it e.g.} one can choose $ n^N $ to be real).

	To obtain the effective sigma model on the string  world sheet \cite{ABEKY,SYmon,GSY05}, 
	one assumes the moduli $ n^l $ to be slowly-varying functions along the string, $ n^l ~=~ n^l(x^k) $.
	Substituting then the solution \eqref{str} into the kinetic terms of the action \eqref{theory}, one
	arrives at the CP($N-1$) sigma model (see the review \cite{SYrev} for details)
\begin{align}
\label{cp}
	S_{\text{CP($N-1$)}}^{1+1} ~~=~~ 2\beta\, \int\, dt\,dz 
					\Bigl\{\, \left|\p n^l\right|^2    
						  ~+~  \left(\nbar \p_k n\right)^2\,
					\Bigr\}\,.
\end{align}
	Here $ \beta $ is the two-dimensional coupling constant which is obtained from an integral over
	the profile functions of the quark and gauge fields over the transverse plane. 
	Using the first-order differential equations \eqref{foes} one can show that the integral
	is in fact a total derivative, and thus, determined by the boundary conditions \eqref{boundary}.
	This yields 
\beq
\label{beta}
	\beta ~~=~~ \frac{2\pi}{g_2^2}\,.
\eeq
	In quantum theory both coupling constants entering this equation run, and so one has 
	to specify the scale at which the above relation holds.
	It is natural to set the scale of Eq.~\eqref{beta} to the cut-off scale of world-sheet dynamics,
	which is given by the inverse thickness of the string $ g_2 \sqrt{\xi} $.

	Below $ g_2\sqrt{\xi} $ the four-dimensional gauge couplings do not run due to the breaking
	of the gauge symmetry. 
	The two-dimensional coupling starts logarithmic run below the cut-off scale,
\beq
\label{asyfree}
	4 \pi \beta ~~=~~ N\,  \ln \lgr \frac{E}{\Lambda_\text{CP($N-1$)}} \rgr.
\eeq
	By itself the CP($N-1$) theory is asymptotically free \cite{P75}.

	In the limit of large $ \mu $, the bulk theory becomes \none SQCD with its own scale \eqref{Lambda}.
	Using Eqs.~\eqref{beta} and \eqref{asyfree}, one can find the relation between the scales
	of the world-sheet and bulk theories \cite{GSYmmodel}, 
\beq
\label{cpscale}
	\Lambda_\text{CP($N-1$)} ~~=~~ \frac{ \Lambda_{\mc{N}=1}^2 } { g_2\, \sqrt{\xi}}\,,
\eeq
	where the coupling constant is determined by the scale $ \Lambda_{\mc{N}=1} $.

	For a 1/2-BPS string, Eq.~\eqref{cp} gives only half of the world-sheet action, {\it i.e.} the bosonic part.
	The fermionic part of the theory is related to the bosonic one by supersymmetry.
	For the string in the \ntwo microscopic theory at hand, world-sheet dynamics is given 
	by \ntwot CP($N-1$) sigma model,
\begin{align}
\notag
\mc{S}_{\rm 1+1}^{\rm (2,2)}  ~~=~~ 2\beta
	\int  d^2x
	\biggl\lgroup\; 
	&
	\left|\p_k n \right|^2  ~+~ \left(\ov{n}\p_k n\right)^2  
	~+~ \ov{\xi}{}_L\, i\p_R\, \xi_L  ~+~ \ov{\xi}{}_R\, i\p_L\,  \xi_R 
	\\[2mm]
\label{str_ntwot}
	&\;
	~~-~
	i \left(\nbar\p_R n\right)\, \bxil\xil ~-~ i \left(\nbar\p_Ln\right) \, \bxir\xir 
	\\[2mm]
\notag
	&\;
		~~+~
		\bxil \xir \bxir \xil ~-~ \bxir \xir \bxil \xil
	\biggr\rgroup ,
\end{align}
	where $ \xi^i $ is the two-dimensional fermionic superpartner of the orientational moduli $ n^l $.
	The translational sector is completely decoupled from dynamics encoded in  Eq.~\eqref{str_ntwot}.

	When \ntwo supersymmetry is broken, the string internal dynamics is altered, and, as was shown
	in \cite{Edalati}, the world sheet theory is given by \ntwoo \CPCn.
	This theory has one dimensionless coupling $ \tgamma $
\footnote{
In this paper the heterotic deformation parameter $ \tgamma $ is related to the analogous parameter
$ \gamma $ introduced in \cite{SYhet} as 
\[
	\tgamma ~~=~~ \sqrt{2/\beta}\,\gamma\,.
\]}
 \cite{SYhet,BSYhet}, which is determined by the
	measure of supersymmetry breaking.
	In two-dimensional theory, this parameter sets the strength of the coupling of the translational
	sector to the orientational one,
\begin{align}
\notag
S_{1+1}^{(0,2)} ~~=~~ 2\beta
	\int & d^2 x 
\lgr
	\bzr\, i\p_L\, \zr ~~+~~ \dots 
\right.
	\\[2mm]
\notag
	&
	\;\;
	+~~
	\left|\p n\right|^2 ~~+~~ \left(\ov{n}\p_k n\right)^2 ~~+~~
	\bxir \, i\p_L \, \xir  ~~+~~ \bxil \, i\p_R \, \xil 
	\\[2mm]
\label{world02}
	&
	\;\;
	-~~
	i \left(\ov{n}\p_L n\right)\, \bxir \xir ~~-~~ 
	i \left(\ov{n}\p_R n\right)\, \bxil \xil  
	\\[2mm]
\notag
	&
	\;\;
	+~~
	\tgamma\, (i\p_L\nbar) \xir\zr ~~+~~ \ov{\tgamma}\, \bxir (i\p_L n) \bzr ~~+~~
	|\tgamma|^2\, \bxil\xil \bzr\zr  
	\\[2mm]
\notag
	&
	\;\;
\left.
	+~~ 
	\left( 1 \;-\; |\tgamma|^2 \right)\, \bxil\xir \bxir\xil  
	~~-~~ \bxil\xil \bxir\xir
\rgr .
\end{align}
	The ellipses denote the left-handed part of the translational sector, which stays decoupled.

	The world sheet theory \eqref{world02} can be obtained from the gauged formulation of CP($N-1$) \cite{W93}.
	Details of the derivation of the \CPC action are given in \cite{SYhet,BSYhet}.
	In this formulation, the most natural parameter of the theory $\delta$ arises as 
	a constant in the quadratic deformation of the superpotential,
\[
	\mathcal{W}_{1+1} ~~=~~ \frac{1}{2}\,\delta\,\Sigma^2\,,
\]
	where $\Sigma$ is a chiral superfield, a part of the gauge supermultiplet.
The parameter $ \delta $ is related to $ \tgamma $ via
\[
	\tgamma ~~=~~ \frac { \sqrt{2}\,\delta } { \sqrt{ 1 +  2 |\delta|^2 } }\,.
\]
	The gauged formulation has a somewhat more direct physical interpretation than
	the representation \eqref{world02}, as the quantum behavior of the system is more directly
	seen in this picture \cite{SYhet2}. 
	In this sense, $ \delta $ is also a more physical parameter, {\it e.g.}
	$ \delta \to \infty $ supposedly corresponds to a conformal phase of the world sheet theory.

	For a non-Abelian string in \ntwo SQCD broken down to \none by soft mass terms $ \mu_1 $ and $ \mu_2 $
	(see Eq.~\eqref{none}), the relation between $ \tgamma $ and $ \mu $ was found to be logarithmic \cite{SYhet,BSYhet}
\beq
	\delta ~~=~~ 
	{\rm const} \cdot \sqrt{\ln\, \frac{g_2^2\mu}{m_W}}
\eeq
	for large $ \mu $.
	The large logarithm is associated with the emergence  of the Higgs branch
	with light particles in \none SQCD in the limit of heavy  adjoint superfields.

	When the $ M $ field is present, the Higgs branch does not develop in the large-$ \mu $
	limit, and the adjoint
	fields are safely integrated out without disrupting the \ntwoo CP($N-1$) theory.
	The two-dimensional parameter $ \tgamma $ is then determined by the microscopic parameter $ h $ of the  $ M $ model. 
	Below we find the fermionic zero modes in the vortex background in the $M$ model,
	and use them to obtain the relation between these two parameters. 
	
%
%
\section{Derivation of the Fermionic Part of the World-sheet Theory}
\setcounter{equation}{0}

	The fermionic part of the $M$ model \eqref{mmodel} is\footnote{
We denote the fermionic superpartner of the $M$ field as $\vartheta$, in contrast to \cite{GSYmmodel},
where $\zeta$ was used. Here $\zeta$ is reserved for the world-sheet supertranslational variable.}
\begin{align}
\notag
	\mc{L}_\text{$M$-model}^\text{ferm} & ~~=~~ 
		\frac{2i}{g_2^2}\, \Tr\, \ov{\lambda}{}^\text{SU($N$)} \ov{\slashed{\md}} \lambda^{\text{SU($N$)}}
		~+~ \frac{4i}{g_1^2}\, \ov{\lambda}{}^\text{U(1)} \ov{\slashed{\p}} \lambda^\text{U(1)}
	\\[3mm]
\notag
	&~
 		~+~ \Tr\, i\, \ov{\psi \slashed{\md}} \psi  
		~+~ \Tr\, i\, \wt{\psi} \slashed{\md} \ov{\wt{\psi}}
		~+~ \frac{2i}{h}\, \Tr\, \ov{\vartheta\,\slashed{\p}}\,\vartheta
	\\[3mm]
\label{fermact}
	&~
		~+~
		i \sqrt{2}\, \Tr 
		\lgr \ov{q}\,\lambda^\text{U(1)}\psi ~-~ \wt{\psi}\,\lambda^\text{U(1)}\ov{\wt{q}}
		 ~+~ \ov{\psi\, \lambda}{}^\text{U(1)} q ~-~ \wt{q}\,\ov{\lambda}{}^\text{U(1)}\ov{\wt{\psi}} \rgr
	\\[3mm]
\notag
	&~
		~+~
		i \sqrt{2}\, \Tr
		\lgr \ov{q}\,\lambda^\text{SU($N$)}\psi ~-~ \wt{\psi}\,\lambda^\text{SU($N$)}\ov{\wt{q}}
		 ~+~ \ov{\psi\, \lambda}{}^\text{SU($N$)}q ~-~ \wt{q}\,\ov{\lambda}{}^\text{SU($N$)}\ov{\wt{\psi}} \rgr
	\\[3mm]
\notag
	&~
		~+~
		i\,\Tr \lgr
			    \wt{q}\, \psi\,\vartheta ~+~ \wt{\psi}\,q\,\vartheta 
			~+~ \ov{\psi\, \wt{q}\,\vartheta}  ~+~ \ov{q\,\wt{\psi}\,\vartheta} \rgr
	\\[3mm]
\notag
	&~
		~+~
		i\,\Tr \lgr
				\wt{\psi}\,\psi\,M ~+~ \ov{\psi\, \wt{\psi}\, M} \rgr ,
\end{align}
where $\lambda^{\alpha}$'s are fermionic \none superpartners of gauge fields, while $\psi^{\alpha}$,
 $\tilde{\psi}_{\dot{\alpha}}$ are matter fermions, $\alpha$, $\dot{\alpha}=1,2$ are 
 their spinor indices.
	We use the decomposition
\[
	\vartheta^A_B ~~=~~ \frac{1}{2}\,\delta^A_B\, \vartheta^0  ~+~
				(T^a)^A_B\, \vartheta^{a}  ~~\equiv~~
			\frac{1}{2}\,\delta^A_B\, \vartheta^0  ~+~
				(\vartheta^N)^A_B
\]
	for the fermionic superpartner of the $M$ field.
	We need to find the fermionic zero modes in the background of the vortex string \eqref{str}.

	In \cite{GSYmmodel} an index theorem was derived, which shows that this theory possesses
	$ 4 $ $ + $ $ 4 ( N - 1 ) $ zero modes.
	The first four correspond to the fermionic superpartners $ \zeta $ of the bosonic translational moduli
	$ x_0^1 $ and $ x_0^2 $ of the world-sheet theory. 
	The other $ 4 ( N - 1 ) $ are the superorientational modes which are associated with the 
	fermionic superpartners $ \xi^l $ of the orientational moduli $ n^l $.

	Since the $M$ model possesses 1/2 of  supersymmetry of the original \ntwo theory, one can utilize
	it in order to find one half of the fermionic zero modes --- the ones that are associated 
	with the left-handed fermions of the string world sheet.
	These supertransformations are identical to those of the original \ntwo theory, and the corresponding
	zero modes were calculated in \cite{BSYhet}.

	We have, for the supertranslational modes,
\begin{align}
\label{N2_strans}
\notag
\ov{\psi}_{\dot{2}}	& ~~=~~  -\,  2\sqrt{2}\, \frac{x_1 ~+~ i x_2}{N r^2} \,
		\lgr \frac{1}{N} \phi_1 ( f + (N-1) f_N ) ~+~ \frac{N-1}{N} \phi_2 ( f - f_N )  \right.
		\\[2mm]
\notag
			& \phantom{~~=~~  -\,  2\sqrt{2}}
			~+~ \left( n\nbar ~-~ 1/N \right )
			\Bigl\{ \phi_1 ( f + ( N-1 ) f_N ) ~-~ \phi_2 ( f - f_N) \Bigr\}
		\left. \rgr\, \zeta_L \,,
		\\[2mm]
\lambda^{1\ \rm U(1)} 	& ~~=~~ -\, \frac{i g_1^2}{2} \lgr (N-1)\phi_2^2  ~+~ \phi_1^2 ~-~ N\xi \rgr \, \zeta_L \,,
		\\[2mm]
\notag
\lambda^{1\ \text{SU($N$)}}	& ~~=~~ -\, {i g_2^2}\, ( n\nbar ~-~ 1/N )\, \lgr \phi_1^2 ~-~ \phi_2^2 \rgr\, \zeta_L\,.
\end{align}
At the same time, for the superorientational modes we have
\begin{align}
\label{N2_sorient}
\notag
\overline{\psi}_{\dot{2}Ak} & ~~=~~ \frac{\phi_1^2 ~-~ \phi_2^2}{\phi_2} \cdot n \overline{\xi}_L  \,,
 \\[2mm]
\lambda^{1\ \text{SU($N$)}} & ~~=~~ i \sqrt{2}\, \frac{ x^1 ~-~ i\, x^2 }{r^2} 
						  \frac{\phi_1}{\phi_2} f_N \cdot n \overline{\xi}_L \,.
\end{align}
	Note that in Eqs.~\eqref{N2_strans} and \eqref{N2_sorient} we listed only  nonvanishing 
	components, which are (by definition) proportional to the left-handed fermions. 
	As shown in \cite{GSYmmodel}, the $M$ model has a U(1)$_{\wt R}$ symmetry, under which these components
	have the charge +1. 
	There must exist other, right-handed zero modes which are positively charged 
	under the U(1)$_{\wt R}$ symmetry  as well.
	Although we will not need explicit expressions for the 
	zero modes \eqref{N2_strans}, \eqref{N2_sorient} in what follows,
	they are helpful in finding a good {\it ansatz} for the right-handed zero modes.
	It is rather obvious that by substituting the modes \eqref{N2_strans} and \eqref{N2_sorient}
	into the kinetic terms of Eq.~\eqref{fermact} one recovers the left-handed kinetic part of the 
	\CPC model \eqref{world02}.

	To obtain the right-handed zero modes one generally needs to solve the Dirac equations.
	We follow the approach of \cite{GSYmmodel} where the parameter $ h $ was tuned to be  
	small (but non-zero),
\beq
\label{smallh}
	0 ~~<~~ h ~~\ll~~ g_2^2\,,
\eeq
	allowing to find the solution analytically.
	We deal with supertranslational and superorientational modes in turn.

	First we make a guess on what fields should participate in the right-handed modes. 
	From the mass-deformed \ntwo case \cite{BSYhet} we know that they would involve
	the fields $ \ov{\wt{\psi}}{}_{\dot 1} $ and $ \lambda^{22} $ but
	the latter field is not present in our $M$ model as it was integrated out.
	We infer then that the correct set of fermions which constitute the right-handed modes
	are $ \ov{\wt{\psi}}{}_{\dot 1} $, $ \vartheta^0 $ and $ \vartheta^a $, {\it i.e.}
	those which have the U(1)$_{\wt R}$ charges +1 and couple to $ \ov{\wt{\psi}}{}_{\dot 1} $.
	The fermions $ \lambda^{1} $ and $\bar{\psi}_{\dot{2}}$
        decouple from this set completely. They are given by (\ref{N2_strans}), (\ref{N2_sorient}).

\subsection{Supertranslational Zero Modes}

	The Dirac equations for the $ \ov{\wt \psi}{}_{\dot 1} $ and $ \vartheta $ are
\begin{align*}
&
	i\, \slashed{\nabla}^{2\dot{1}}\, \ov{\wt\psi}{}_{\dot 1}  
		~+~  i\, q \lgr \frac{1}{2}\, \vartheta_0  ~+~ \vartheta^N \rgr ~~=~~ 0\,, \\
&
	\frac{i}{h}\, \ov{\slashed{\p}}{}_{\dot{1}2}\, (\vartheta^N)^2
		~+~ \frac{i}{2}\, {\rm Traceless} \left\{ \ov{q \wt{\psi}}{}_{\dot 1} \right\} ~~=~~ 0\,, \\
&
	\frac{i}{h}\, \ov{\slashed{\p}}_{\dot{1}2}\, \vartheta^0 
		~+~ \frac{i}{N}\, {\rm Tr} \lgr \ov{q \wt{\psi}}{}_{\dot 1} \rgr ~~=~~ 0\,.
\end{align*}

For constructing an appropriate {\it ansatz} for the solution we split the trace and traceless components of the
fields, which we mark by superscripts $0$ and $N$, respectively, attributing different profile functions to them. 
One has a freedom of placing a factor of $ (x^1 \pm ix^2)/r $ in these components, which gives two possibilities
for the {\it ans\"atze} for the zero modes.
Let us call $ \zr $ and $ \bzr $ the corresponding world-sheet fermions; then 
\begin{align*}
	\ov{\wt{\psi}}{}_{\dot 1} ~~=~~ & \frac{1}{2}\, \lgr \cts ~+~ N (n\nbar \;-\; 1/N)\,\ctN \rgr \zr ~+~ \\
	& 
		\quad~+~
		\frac{1}{2}\, \frac{x^1 \,-\, ix^2}{r}\, \lgr \pts ~+~ N(n\nbar \;-\; 1/N)\, \ptN \rgr \bzr\,, \\
	(\vartheta^0)^2 ~~=~~ &
		\frac{x^1 \,+\, ix^2}{r}\, \rts(r)\cdot \zr 
		~+~ \tts(r)\cdot \bzr\,, \\
	(\vartheta^N)^2 ~~=~~ &
		\frac{x^1 \,+\, ix^2}{r}\,\rtN(r) (n\nbar \;-\; 1/N) \cdot \zr 
		~+~ \ttN(r)\cdot (n\nbar \;-\; 1/N) \cdot \bzr \,.
\end{align*}
Here the superscript ``tr'' is used to denote the profile functions of the supertranslational modes, versus the
superorientational modes to appear later.
Substituting this into the Dirac equations, one obtains the equations 
for the profile functions 
$ \chi_{\rm tr}^{0,N} $ and $ \rho_{\rm tr}^{0,N} $,
\begin{align*}
	&
	\p_r\, \cts ~-~ \frac{1}{Nr}\lgr f\,\cts ~+~ (N-1)\, f_N\, \ctN \rgr \\
	&
	\qquad\qquad
		~+~ i\, \frac{ \phi_1 ~+~ (N-1)\,\phi_2}{N} \rts 
		~+~ 2i\, \frac{N-1}{N^2}\, (\phi_1 ~-~ \phi_2)\, \rtN ~~=~~ 0\,,
\\
	&
	\p_r\, \ctN ~-~ \frac{1}{Nr} \lgr f\, \ctN ~+~ f_N\, \cts ~+~ (N-2)\,f_N\,\ctN \rgr \\
	& 
	\qquad\qquad\qquad
		~+~ i\, \frac{\phi_1 ~-~ \phi_2}{N}\, \rts 
		~+~ 2i\, \frac{(N-1)\phi_1 ~+~ \phi_2}{N^2}\, \rtN ~~=~~ 0\,,
\\
	&
	-\,\frac{1}{h} \left\{ \p_r ~+~ \frac{1}{r} \right\} \rtN 
		~+~ \frac{i}{4} \lgr (\phi_1 ~+~ (N-1)\,\phi_2)\,\ctN \right. \\
	&
	\qquad\qquad
		\left.
		~+~ (\phi_1~-~\phi_2)\,\cts ~+~ (N-2)\,(\phi_1~-~\phi_2)\,\ctN \rgr ~~=~~ 0\,,
\\
	&
	-\,\frac{1}{h} \left\{ \p_r ~+~ \frac{1}{r} \right\} \rts 
		~+~ \frac{i}{2N} \lgr (\phi_1 ~+~ (N-1)\,\phi_2)\,\cts \right. \\
	&
	\qquad\qquad\qquad\qquad\qquad\qquad\quad
		\left.
		~+~ (N-1)\,(\phi_1~-~\phi_2)\,\ctN \rgr ~~=~~0\,,
\end{align*}
	and 
\begin{align}
\label{str_eqn}
\notag
	&
	\left\{ \p_r ~+~ \frac{1}{r} \right\} \pts ~-~ \frac{1}{Nr}\lgr f\,\pts ~+~ (N-1)\, f_N\, \ptN \rgr \\
\notag
	&
	\qquad\qquad
		~+~ i\, \frac{ \phi_1 ~+~ (N-1)\,\phi_2}{N} \tts 
		~+~ 2i\, \frac{N-1}{N^2}\, (\phi_1 ~-~ \phi_2)\, \ttN ~~=~~ 0\,,
\\
\notag
	&
	\left\{ \p_r ~+~ \frac{1}{r} \right\} \ptN ~-~ \frac{1}{Nr} \lgr f\, \ptN ~+~ f_N\, \pts ~+~ (N-2)\,f_N\,\ptN \rgr \\
	& 
	\qquad\qquad\qquad
		~+~ i\, \frac{\phi_1 ~-~ \phi_2}{N}\, \tts 
		~+~ 2i\, \frac{(N-1)\phi_1 ~+~ \phi_2}{N^2}\, \ttN ~~=~~ 0\,,
\\
\notag
	&
	-\,\frac{1}{h}\, \p_r\, \ttN 
		~+~ \frac{i}{4} \lgr (\phi_1 ~+~ (N-1)\,\phi_2)\,\ptN \right. \\
\notag
	&
	\qquad\qquad
		\left.
		~+~ (\phi_1~-~\phi_2)\,\pts ~+~ (N-2)\,(\phi_1~-~\phi_2)\,\ptN \rgr ~~=~~ 0\,,
\\
\notag
	&
	-\,\frac{1}{h}\, \p_r\, \tts 
		~+~ \frac{i}{2N} \lgr (\phi_1 ~+~ (N-1)\,\phi_2)\,\pts \right. \\
\notag
	&
	\qquad\qquad\qquad\qquad\qquad\qquad\quad
		\left.
		~+~ (N-1)\,(\phi_1~-~\phi_2)\,\ptN \rgr ~~=~~0
\end{align}
	for the profile functions $ \psi_{\rm tr}^{U,N} $ and $ \vartheta_{\rm tr}^{0,N} $.
	Only the second set of equations turns out to yield solutions finite at $ r \to 0 $.
	So we drop $ \chi_{\rm tr} $ and $ \rho_{\rm tr} $, and accept
\begin{align}
\notag
	(\vartheta^0)^2 & ~~=~~ \tts(r) \cdot \zr\,,
\\
\label{tr_anz}
	(\vartheta^N)^2 & ~~=~~ \ttN(r)\, (n\nbar ~-~ 1/N) \cdot \zr\,,
\\
\notag
	\ov{\wt{\psi}}{}_{\dot 1} & ~~=~~ \frac{1}{2}\, \frac{x^1 \,-\, ix^2}{r} 
						\lgr \pts ~+~ N\,(n\nbar ~-~ 1/N)\, \ptN \rgr \, \zr
\end{align}
	for the zero modes.
	Following \cite{GSYmmodel,BSYhet} we solve Eqs.~\eqref{str_eqn} in the limits of large $r$, {\em i.e.}
	$ r \gg 1/(g_2\sqrt{\xi}) $ and intermediate $r$, {\em i.e.} $ r \lesssim 1/(g_2\sqrt{\xi}) $.
 The idea is that if $h$ is small (see (\ref{smallh})) we have the  matter fields in our theory
which are much lighter than the gauge bosons, see (\ref{U1mass}), (\ref{SUNmass}). 
Although the light scalar fileds vanish on the string solution, 
the presence of light fermion fields affects the fermionic sector of the theory \cite{SYnone,SYhet}.
In particular, the string fermion zero modes have a two-layer structure in the plane
orthogonal to the string axis: a core of the size of the inverse gauge boson mass plus long-range
``tails" formed by light fermions.
	Completely analogously to calculations in \cite{GSYmmodel,BSYhet}, in the limit of 
	small $h$, we find in the large-$r$ domain,
\begin{align*}
	&& \tts &~~=~~ -\, C\,i\,\frac{m_0^2}{\sqrt{\xi}}\, K_0(m_0 r) \,,
	& \ttN & ~~=~~ -\, C\,i\,\frac{N}{2}\,\frac{m_0^2}{\sqrt{\xi}}\,K_0(m_0 r)\,, \\
	&& \pts &~~=~~ -\, C\,\p_r\,K_0(m_0 r) \,,
	& \ptN & ~~=~~ -\, C\,\p_r\,K_0(m_0 r)\,,
\end{align*}
	where $ K_0(z) $ is the McDonald function, 
	while at intermediate $ r $ we get
\begin{align*}
	\pts & ~~=~~ \ptN ~~=~~ \frac{C}{\sqrt{\xi}}\,\frac{\phi_1}{r}\,,
	\\
	\tts & ~~\simeq~~ -\,C\,i\,\frac{m_0^2}{\sqrt{\xi}}\,\ln\frac{m_W}{m_0}\,,
	\\
	\ttN & ~~\simeq~~ -\,C\,i\,\frac{N}{2}\,\frac{m_0^2}{\sqrt{\xi}}\,\ln\frac{m_W}{m_0}\,,
\end{align*}
where 
\[
	m_0 ~~\equiv~~ \sqrt{\frac{h}{2}\, \xi}\,.
\]
	The arbitrary constant $ C $ is common for all profile functions, and we can safely put
	 $ C = 1 $.

\subsection{Superorientational Zero Modes}

Orientational fermion zero modes in the $M$ model with the gauge group U(2) were calculated in
\cite{GSYmmodel}. Here we generalize these results to the theory with the U($N$) gauge group.	
Now the trace component of the $ \vartheta $ field  is not involved; therefore, we
	need to deal only with two Dirac equations,
\begin{align}
\label{sor_dirac}
&
\notag
	i\, \slashed{\nabla}^{2\dot{1}}\, \ov{\wt\psi}{}_{\dot 1}  
		~+~  i\, q\, \vartheta^N  ~~=~~ 0\,, \\
&
	\frac{i}{h}\, \ov{\slashed{\p}}{}_{\dot{1}2}\, (\vartheta^N)^2
		~+~ \frac{i}{2}\, {\rm Traceless} \left\{ \ov{q \wt{\psi}}{}_{\dot 1} \right\} ~~=~~ 0\,.
\end{align}
	The form of the zero modes \eqref{N2_sorient} prompts  us that the right-handed modes may be proportional
	to $ n\bxir $, or $ \xir\nbar $, which gives two possibilities for the {\it ansatz}.
	One also has the freedom of putting the factor of $ (x^1 \pm ix^2)/r $ in either 
	$ \ov{\wt \psi}{}_{\dot 1} $ or  $ \vartheta^N $.
	Overall, we write the following four {\it ans\"atze}:
\begin{align*}
	(\vartheta^N)^2  & ~~=~~ 2\,\tor(r)\cdot n\bxir ~+~  2\,\eor(r) \cdot \xir \nbar \\
			& \quad
			~+~ 2\,\frac{x^1\,+\,ix^2}{r}\, \kor(r) \cdot n\bxir 
			~+~ 2\,\frac{x^1\,+\,ix^2}{r}\, \sor(r) \cdot \xir\nbar \,,\\
	\ov{\wt{\psi}}{}_{\dot 1} & ~~=~~ 2\, \frac{x^1\,-\,ix^2}{r}\, \por(r)\cdot n\bxir 
				      ~+~ 2\, \frac{x^1\,-\,ix^2}{r}\, \cor(r)\cdot \xir\nbar \\
				& \quad
				          ~+~ 2\, \uor(r)\cdot n\bxir
					  ~+~ 2\, \oor(r)\cdot \xir\nbar\,,
\end{align*}
	where the subscript ``or'' is added to distinguish the profile
	functions from those of the translational modes. 
	Plugging these into the Dirac equations \eqref{sor_dirac}, we have eight equations for the profile
	functions
\begin{align}
\label{sor_eqn}
\notag
	&
	-\, \p_r\, \tor(r) ~+~ \frac{ih}{2}\, \phi_1(r)\,\por(r) ~~=~~ 0\,,  \\
\notag
	&
	\phantom{-\,}
	\left\{ \p_r ~+~ \frac{1}{r} \right\}\, \por(r)  
			~-~ \frac{1}{Nr}\, \lgr f ~+~ (N-1)\,f_N \rgr \por(r) 
			~+~ i\,\phi_1(r)\,\tor(r) ~~=~~ 0   \,,              \\[4mm]
\notag
	& 
	-\, \left\{ \p_r ~+~ \frac{1}{r} \right\}\, \eor(r) 
			~+~ \frac{ih}{2}\, \phi_1(r)\,\cor(r) ~~=~~ 0\,, \\
	&
	\phantom{-\,}
	\p_r\, \cor(r) ~-~ \frac{1}{Nr}\, \lgr f ~+~ (N-1)\,f_N \rgr \cor(r) 
			~+~ i\, \phi_1(r)\, \eor(r) ~~=~~ 0 \,,\\[4mm]
\notag
	&
	-\, \p_r\, \kor(r) ~+~ \frac{ih}{2}\, \phi_2(r)\, \uor(r) ~~=~~ 0\,, \\
\notag
	&
	\phantom{-\,}
	\left\{ \p_r ~+~ \frac{1}{r} \right\}\, \uor(r) 
			~-~ \frac{1}{Nr} \lgr f ~-~ f_N \rgr \uor(r) 
			~+~ i\,\phi_2(r)\, \kor(r) ~~=~~ 0 \,,\\[4mm]
\notag
	&
	-\, \left\{ \p_r ~+~ \frac{1}{r} \right\}\, \sor(r) 
			~+~ \frac{ih}{2}\, \phi_2(r)\,\oor(r) ~~=~~ 0 \,,\\
\notag
	&
	\phantom{-\,}
	\p_r\, \oor(r) ~-~ \frac{1}{Nr}\lgr f ~-~ f_N \rgr \oor(r) 
			~+~ i\,\phi_2(r)\,\sor(r) ~~=~~ 0\,.
\end{align}
From this set of four pairs of equations only the first pair yields nonsingular profile functions
for the zero modes.
Thus we have
\begin{align}
\notag
	(\vartheta^N)^2  & ~~=~~ 2\,\tor(r)\cdot n\bxir \,,\\
\label{or_anz}
	\ov{\wt{\psi}}{}_{\dot 1} & ~~=~~ 2\, \frac{x^1\,-\,ix^2}{r}\, \por(r)\cdot n\bxir \,.
\end{align}
Again, we solve the equations \eqref{sor_eqn} separately in the domain of large $ r $ and 
intermediate $ r $, assuming $ h $ to be small.
Parallelizing  \cite{GSYmmodel,BSYhet}, we get for large $ r $
\begin{align}
\label{sor_large}
\notag
	\tor(r) & ~~=~~ -\, \frac{ih\sqrt{\xi}}{2}\, K_0(m_0 r) \,,\\
	\por(r) & ~~=~~ -\, \p_r\, K_0(m_0 r) \,,
\end{align}
while in the intermediate-$r$ domain the profile functions take the form
\begin{align}
\label{sor_interm}
\notag
	\tor(r) & ~~\simeq~~ -\, \frac{ih\sqrt{\xi}}{2}\, \ln \frac{m_W}{m_0} \,,\\
	\por(r) & ~~\simeq~~ \frac{\phi_1}{\sqrt{\xi}\, r}\,.
\end{align}
Relative normalization of equations \eqref{sor_large} and \eqref{sor_interm} has been taken care of
to ensure agreement between the two domains. 
The overall normalization, similarly to the supertranslational case, is given by a common 
arbitrary constant which we have put to one.

We observe that the right-handed zero modes exhibit the long-range $1/r$ behavior similar to that observed
in the \ntwo theory deformed solely by the $\mu_{1,2}$ parameters \cite{GSYmmodel,SYhet}. 
This is expected, as in the limit $ h \to 0 $ the theory re-acquires the Higgs branch, and the associated
massless modes. 
We have no need, however, of taking this limit; we chose $ h $ to be small, 
see Eq.~\eqref{smallh}, only for the purpose of making analytical computations simpler. At the same time
$h$ can be (and is) treated as a fixed parameter. Decoupling of the adjoint fields does not depend
on the value of $h$.

\subsection{Bifermionic Coupling}

The easiest way to obtain the coupling constant $ \tgamma $ of the heterotic ${\rm CP}(N-1)$ model
is to calculate the strength of the coupling of the supertranslational and superorientational
modes induced on the string world sheet
\beq
\label{bif_norm}
	\mc{L}_\text{eff}^{\mc{N}=(0,2)} ~~\supset~~
	2\beta \cdot I_{\zeta\xi}\, ( i\p_L\nbar\,\xir\zr ~+~ \bxir\,i\p_L n\, \bzr )\,,
\eeq
where we separate the factor  $ 2\beta $ for convenience. 
It is natural to assume that $ \tgamma $ will be real, since the deformation $h$ is.

To be able to compare this coupling constant to $ \tgamma $ in Eq.~\eqref{world02} one has to normalize 
the participating fermions.
We define the normalization integrals $ I_\zeta $ and $ I_\xi $ for the fermions $ \zeta $ and $ \xi $ as
\beq
\label{kin_norm}
	\mc{L}_\text{eff}^{\mc{N}=(0,2)} ~~\supset~~
	2\beta \cdot ( I_\zeta\, \bzr\, i\p_L\,\zr  ~+~ I_\xi\, \bxir\, i\p_L \xir ) \,.
\eeq
Substituting the expressions \eqref{tr_anz} and \eqref{or_anz} for the zero modes into the kinetic terms
of the microscopic theory \eqref{fermact} one obtains
\begin{align*}
	I_\zeta & ~~=~~ \frac{N}{2\xi}\, \frac{m_W^2}{4}\, \int\, r\,dr 
		\biggl\lgroup -\, \frac{2}{h}\, 
			  \left\{ \left( \tts \right)^2 
				~+~ 4\,\frac{N-1}{N^2}\, \left( \ttN \right)^2 \right\} \\
		& \qquad\qquad\qquad\qquad\qquad~~
			~+~ \left( \pts \right)^2 ~+~ (N-1)\,\left( \ptN \right)^2  \biggr\rgroup\,, \\
	I_\xi & ~~=~~ 2\,g_2^2\, \int\, r\,dr 
		\lgr -\,\frac{2}{h}\bigl( \tor(r) \bigr)^2 ~+~ \bigl( \por(r) \bigr)^2 \rgr.
\end{align*}
From a similar procedure one extracts the expression for the bifermionic coupling
\[
	I_{\zeta\xi} ~~=~~ \frac{g_2^2}{2}\, \int\, r\,dr
		\lgr -\,\frac{4}{h}\,\tor\,\ttN  ~+~ N\, \por\,\ptN  
			~+~ \rho\, \por \left( \pts ~-~ \ptN \right) \rgr.
\]
Substituting here the profile functions of the zero modes, in particular, their $ 1/r $-tails,
we obtain with logarithmic accuracy
\begin{align*}
	I_\zeta & ~~=~~ N^2\,\frac{g_2^2}{8}\, \ln \frac{m_W}{m_0}\,, \\
	I_\xi   & ~~=~~ 2\,g_2^2\, \ln \frac{m_W}{m_0}\,, \\
	I_{\zeta\xi} & ~~=~~ N\, \frac{g_2^2}{2}\,  \ln \frac{m_W}{m_0}\,.
\end{align*}
The logarithms come from the domain $ 1/m_W \ll r \ll 1/m_0 $  of the zero mode profile
functions.

Normalizing the world-sheet fermions using the above integrals, one arrives at the answer
for the world-sheet coupling $ \tgamma $ up to a contribution suppressed by the inverse large logarithms.
Parametrically, the logarithms under consideration can be written as 
\[
	\ln \frac{m_W}{m_0}  ~~\simeq~~ 
			\ln \frac{g_2}{\sqrt{h}} ~~=~~ \frac12 \ln \frac{g_2^2}{h}\,,
\]
where $ h $ is small.
Overall our result takes the form
\[
	\tgamma ~~=~~ \frac{\sqrt{2}\delta}{\sqrt{1 ~+~ 2|\delta|^2}} ~~=~~ 
		\frac{I_{\zeta\xi}}{\sqrt{I_\zeta\, I_\xi}}  ~~=~~
			1 ~+~ O\lgr \frac{1}{\ln g_2^2/h } \rgr
\]
	(remind that in this paper $ \tgamma $ is related to $ \gamma $ of \cite{SYhet} as
	$ \tgamma = \sqrt{2/\beta}\, \gamma $).
Therefore, 
\beq
\label{deltaresult}
	\delta ~~=~~ \text{const}\cdot \sqrt{\ln g_2^2/h}\,.
\eeq

This relation is, of course, quite analogous to that for the deformation parameter $ \delta $ 
in the heterotic string scenario of Refs.~\cite{SYhet,BSYhet}. 
The difference, however, is that $ \delta $ does not go  all the way to infinity
in the limit $\mu\to\infty$; hence  the \CPC model 
gives a reliable description of the string world sheet in this limit.

\section{Confined monopoles}
\setcounter{equation}{0}

Since the bulk theory quarks are in the Higgs phase, the monopoles are confined. It was shown in
\cite{Tong,SYmon,HT2} that when we introduce a nonvanishing FI parameter $\xi$  
in U$(N)$ \ntwo SQCD, the   't Hooft--Polyakov monopoles of 
the SU$(N)$ subgroup become confined on the string --- 
they become string junctions of two elementary non-Abelian strings. Each string of the bulk theory corresponds to a particular vacuum of the world-sheet theory. In particular, \ntwot supersymmetric CP$(N-1)$ model on the string world sheet
has $N$ degenerate vacua and kinks interpolating between different vacua. These kinks are interpreted as 
confined monopoles of the bulk theory \cite{Tong,SYmon,HT2}. In the limit of massless quarks
these monopoles become truly non-Abelian. They no longer carry average magnetic flux since
\beq
\langle n^l\rangle =0
\eeq
in the strong coupling limit of the CP$(N-1)$ model. Still these monopoles (= kinks) are stabilized by  quantum effects in the CP$(N-1)$ model. 
They acquire mass and an inverse size of the  order of $\Lambda_{{\rm CP}(N-1)}$. They 
are described by fields $n^l$ and form the fundamental representation of the \cfl group \cite{W79}.

Now what happens with these monopoles when we introduce \ntwo supersymmetry breaking parameter
$\mu$ and tend it to infinity converting the microscopic theory  into \none SQCD? 
(We assume that $\mu_1\sim\mu_2\equiv\mu$).
Note that \none SQCD has no adjoint fields at all (they completely decouple), so in no
way the monopoles can be seen quasiclassically. 
Moreover, no breaking of the gauge group to an Abelian subgroup occurs in this theory;
therefore, the monopoles (if exist) should be truly non-Abelian.

This question was addressed in \cite{SYhet}, where it was noted that \none SQCD  develops 
a Higgs branch in the  limit $\mu\to\infty$, 
and therefore the fate of the confined monopoles can be traced only up to a
 finite value of $\mu$, see Eq.~(\ref{critmu}). On the other hand, in the $M$ model
there is no Higgs branch in the limit $\mu\to\infty$ and the presence of confined monopoles was traced all the way to
 $\mu = \infty$ \cite{GSYmmodel}.
In this limit we get a remarkable result: although the adjoint fields 
are eliminated from our theory 
and  monopoles cannot be seen in any semiclassical description,
our  analysis shows
that confined non-Abelian monopoles still exist in the theory (\ref{mmodel}). They are seen
as ${\rm CP}(N-1)$-model kinks in the effective world-sheet theory on the non-Abelian string.

The only loophole in the above argument is that the fermionic sector of the world-sheet theory
was not studied in \cite{GSYmmodel}. In fact, it was not clear, whether the world-sheet theory has
$N$ strictly degenerate  vacua and kinks interpolating between them (to be interpreted as
confined  monopoles). Say, if $N$ vacua were split (as it happens in the nonsupersymmetric case
\cite{GSY05}) a monopole and antimonopole attached to the string would come close to
each other  to form a 
meson-like configuration, see the  review \cite{SYrev} for details. If in the large-$\mu$ limit the splittings
were large, the binding inside these mesons could become stronger and individual monopoles would not
be seen. This effect corresponds to the kink confinement in two-dimensional 
nonsupersymmetric CP$(N-1)$ model \cite{W79}.

In this paper we completed the proof of the presence of confined  non-Abelian monopoles
in the $M$ model in the limit $\mu\to\infty$. 
By confined we mean confined on the string but unconfined along the string.
Above we demonstrated that the world-sheet
theory on the non-Abelian string is heterotic \ntwoo supersymmetric CP$(N-1)$ model.
We derived Eq.~(\ref{deltaresult}) which relates the deformation parameter of the world sheet theory
to parameters of the bulk theory in the large $\mu$ limit. In particular, it shows that $\delta$
goes to a constant  at large $\mu$.

Physics of the heterotic \ntwoo supersymmetric CP$(N-1)$ model was studied in the large-$N$ approximation in \cite{SYhet2}. In this paper it was shown that supersymmetry is spontaneously broken (see also \cite{Thetdyn}). Still the $Z_{2N}$ discrete symmetry present in the model is
spontaneously broken down to $Z_2$, and the model has $N$ strictly degenerate vacua.
This ensures the presence of kinks, interpolating between these vacua. These kinks are
confined non-Abelian monopoles of the bulk theory.

The kink masses were calculated in \cite{SYhet2} in the large-$N$ approximation. As was already mentioned,
the kinks in the strong coupling regime are described by fields $n^l$ \cite{W79}. Due to spontaneous
supersymmetry breaking the $n^l$ boson masses and those of fermions $\xi$ are different. 
Namely~\cite{SYhet2},
\beq
m_n=\Lambda_{{\rm CP}(N-1)}, \qquad m_{\xi}=\Lambda_{{\rm CP}(N-1)}\,
\exp{\left(-\frac{8\pi\beta}{N}\,|\delta|^2\right)},
\label{kinkmass}
\eeq
where $\delta$ is given in Eq.~(\ref{deltaresult}), while the coupling constant
\beq
\beta=\frac{N}{4\pi}\,\ln{\left(\frac{m_W}{\Lambda_{{\rm SU}(N)}}\right)}.
\eeq
As we see, these expressions
give finite nonvanishing masses for bosonic and fermionic components of  the non-Abelian 
monopoles confined to the string. The masses are well defined in the limit $\mu\to\infty$
since the parameter $\delta$ stays finite in this limit.

\section{Conclusions}
\setcounter{equation}{0}

This paper concludes the program started in \cite{GSYmmodel},
namely direct and explicit derivation of the world-sheet theory  for non-Abelian strings in
the  $M$ model
starting from the bulk theory with \ntwo supersymmetry broken down to \none
by the mass terms of the adjoint fields and the coupling of the quark fields to the $M$ field.
We demonstrated that the-world sheet
theory on the non-Abelian string is heterotic \ntwoo supersymmetric CP$(N-1)$ model.
To this end we had to explicitly obtain all fermion zero modes
 in the limit of  large $\mu $.
We related the  deformation parameter $\delta$ of the world-sheet theory
to parameters of the bulk theory and showed that $\delta$
does not depend on $\mu$ at large $\mu$.

This completes the proof of the presence of non-Abelian monopoles confined to the non-Abelian
strings in the $M$ model. Note that at  $\mu\to\infty$ the bulk theory does not have adjoint fields
and monopoles cannot be seen in the quasiclassical approximation. We showed that they are still
present in the theory and are seen as  kinks in the heterotic  CP$(N-1)$ model on the string. These kinks
are stabilized by nonperturbative effects in two dimensions (e.g. two-dimensional instantons) in the limit
when the color-flavor locked SU$(N)$ symmetry is attained in the bulk theory.

\section*{Acknowledgments}

The work of PAB was supported in part by the NSF Grant No. PHY-0554660. PAB is grateful for kind
hospitality to FTPI, University of Minnesota, where a part of this work was done. 
The work of MS was supported in part by DOE grant DE-FG02-94ER408. 
The work of AY was  supported 
by  FTPI, University of Minnesota, 
by RFBR Grant No. 09-02-00457a 
and by Russian State Grant for 
Scientific Schools RSGSS-11242003.2.

%
%


\small
\begin{thebibliography}{99}
\itemsep -2pt


\bibitem{HT1}
A.~Hanany and D.~Tong,
JHEP {\bf 0307}, 037 (2003)
[hep-th/0306150].

\bibitem{ABEKY}
R.~Auzzi, S.~Bolognesi, J.~Evslin, K.~Konishi and A.~Yung,
Nucl.\ Phys.\ B {\bf 673}, 187 (2003)
[hep-th/0307287].

\bibitem{SYmon}
M.~Shifman and A.~Yung,
Phys.\ Rev.\ D {\bf 70}, 045004 (2004)
[hep-th/0403149].

\bibitem{Tong}
D.~Tong,
Phys.\ Rev.\ D {\bf 69}, 065003 (2004)
[hep-th/0307302].

\bibitem{HT2}
A.~Hanany and D.~Tong,
JHEP {\bf 0404}, 066 (2004)
[hep-th/0403158].

\bibitem{Trev}
D.~Tong,
{\em TASI Lectures on Solitons,}
  arXiv:hep-th/0509216.
 
\bibitem{Jrev}
  M.~Eto, Y.~Isozumi, M.~Nitta, K.~Ohashi and N.~Sakai,
  J.\ Phys.\ A  {\bf 39}, R315 (2006)
  [arXiv:hep-th/0602170].
  
  \bibitem{SYrev}
M.~Shifman and A.~Yung,
{\sl Supersymmetric Solitons,}
Rev.\ Mod.\ Phys. {\bf 79} 1139 (2007)
[arXiv:hep-th/0703267], also an extended version in
{\sl Cambridge University Press, Cambridge, 2009}.


   
\bibitem{Trev2}
D.~Tong,
{\em Quantum Vortex Strings: A Review,}
  arXiv:0809.5060 [hep-th].
  
  \bibitem{Edalati}
  M.~Edalati and D.~Tong,
  JHEP {\bf 0705}, 005 (2007)
  [arXiv:hep-th/0703045].

\bibitem{SYhet}
  M.~Shifman and A.~Yung,
  Phys.\ Rev.\  D {\bf 77}, 125016 (2008)
  [arXiv:0803.0158 [hep-th]].
     
          
     \bibitem{BSYhet}
  P.~A.~Bolokhov, M.~Shifman and A.~Yung,
  [arXiv:0901.4603 [hep-th]].
  
\bibitem{SYhet2}
M.~Shifman and A.~Yung,
  Phys.\ Rev.\  D {\bf 77}, 125017 (2008)
  [arXiv:0803.0698 [hep-th]].


  \bibitem{GSYmmodel}
A.~Gorsky, M.~Shifman and A.~Yung,
Phys.\ Rev.\  D {\bf 75}, 065032 (2007)
  [hep-th/0701040].

\bibitem{HSZ}
A.~Hanany, M.~J.~Strassler and A.~Zaffaroni,
Nucl.\ Phys.\ B {\bf 513}, 87 (1998)
[hep-th/9707244].

\bibitem{VY}
A.~I.~Vainshtein and A.~Yung,
Nucl.\ Phys.\ B {\bf 614}, 3 (2001)
[hep-th/0012250].

\bibitem{GSY05}
A.~Gorsky, M.~Shifman and A.~Yung,
  Phys.\ Rev.\ D {\bf 71}, 045010 (2005)
  [hep-th/0412082].

\bibitem{P75}
A.~M.~Polyakov,
Phys.\ Lett.\ B {\bf 59}, 79 (1975).

\bibitem{W93}
E.~Witten,
  Nucl.\ Phys.\ B {\bf 403}, 159 (1993)
  [hep-th/9301042].



\bibitem{SYnone}
M.~Shifman and A.~Yung,
Phys.\ Rev.\ D {\bf 72}, 085017 (2005)
[hep-th/0501211].

\bibitem{W79}
E.~Witten,
Nucl.\ Phys.\ B {\bf 149}, 285 (1979).

\bibitem{Thetdyn}
  D.~Tong,
  JHEP {\bf 0709}, 022 (2007)
  [arXiv:hep-th/0703235].


\end{thebibliography}
\end{document}